\documentclass[twocolumn,showpacs,preprintnumbers,amsmath,amssymb]{revtex4}
\usepackage{tabularx,graphicx}\begin{document}
%\documentstyle[aps]{revtex}
%\documentstyle[preprint,aps]{revtex}
%\begin{document}
\newcommand{\beq}{\begin{equation}}
\newcommand{\eeq}{\end{equation}}
\newcommand{\beqn}{\begin{eqnarray}}
\newcommand{\eeqn}{\end{eqnarray}}
\newcommand{\bmath}{\begin{subequations}}
\newcommand{\emath}{\end{subequations}}
%\draft
\title{Electrodynamics of superconductors}
\author{J. E. Hirsch }
\address{Department of Physics, University of California, San Diego\\
La Jolla, CA 92093-0319}
 
\date{December 30, 2003} 
\begin{abstract} 
An alternate set of equations to describe the electrodynamics of superconductors at a
macroscopic level is proposed. These equations resemble equations originally
proposed by the London brothers but later discarded by them. Unlike the conventional London equations  the alternate
equations are relativistically covariant, and they can be understood as arising from the 'rigidity' of the
superfluid wave function in a relativistically covariant microscopic theory. They predict that an internal 'spontaneous' electric
field exists in superconductors, and that externally applied electric fields, both longitudinal and transverse,   are screened over a London penetration length, as  magnetic fields are. The associated longitudinal dielectric function predicts a much steeper plasmon dispersion relation than the conventional theory, and a blue shift of the minimum plasmon frequency for
small samples. It is argued that the conventional London equations
lead to difficulties that are removed in the present theory, and that the proposed equations do not contradict any known experimental facts.  Experimental tests are discussed.
\end{abstract}
\pacs{}
\maketitle

\section{Introduction}
It has been generally accepted 
 that the electrodynamics of superconductors in the
'London limit' (where the response to electric and magnetic fields is local)
is described by the London equations\cite{london2,tinkham}. The first London equation
\bmath
\beq
\frac{\partial\vec{J_s}}{\partial t}=\frac{n_s e^2}{m_e} \vec{E}
\eeq
describes the colisionless response of a conducting fluid of density $n_s$,  i.e. free acceleration of the superfluid carriers with charge $e$ and mass $m_e$,  giving rise to the supercurrent $\vec{J}_s$.
The second London equation 
\beq
\vec{\nabla}\times\vec{J}_s=-\frac{n_se^2}{m_e c}\vec{B}
\eeq
\emath
is obtained from Eq. (1a) using Faraday's law and setting a time integration
constant equal to zero, and leads to the Meissner effect.
These equations together with Maxwell's equations
\bmath
\beq
\vec{\nabla}\times\vec{E}=-\frac{1}{c}\frac{\partial\vec{B}}{\partial t}
\eeq
\beq
\vec{\nabla}\times\vec{B}=\frac{4\pi}{c}\vec{J}_s+\frac{1}{c}\frac{\partial\vec{E}}{\partial t}
\eeq
\beq
\vec{\nabla}\cdot\vec{E}=4\pi\rho
\eeq
\beq
\vec{\nabla}\cdot\vec{B}=0
\eeq
\emath
are generally believed to determine the electrodynamic behavior of
superconductors. In this paper we argue that these equations are not correct,
and propose an alternate set of equations. 

There is ample experimental evidence in favor of Eq. (1b), which leads to the
Meissner effect. That equation is in fact preserved in our alternative theory.
However, we argue that there is no experimental evidence for Eq. (1a), even if it
appears compelling on intuitive grounds.
In fact, the London brothers themselves in their early work considered the
possibility that Eq. (1a) may not be valid for superconductors\cite{london1}. 
However, because of the result of an experiment\cite{londonexp} they discarded an alternative possibility and   adopted both
Eqs. (1a) and Eq. (1b), which became known as the 'first' and 
'second' London equations.

It is useful to introduce the electric scalar and magnetic vector potentials
$\phi$ and $\vec{A}$. The magnetic field is given by
\beq
\vec{B}=\vec{\nabla}\times\vec{A}
\eeq
and Eq. (2a) is equivalent to
\beq
\vec{E}=-\vec{\nabla}\phi -\frac{1}{c}\frac{\partial\vec{A}}{\partial t}
\eeq
The magnetic vector potential is undefined to within the gradient of a 
scalar function. The gauge transformation
\bmath
\beq
\vec{A} \rightarrow \vec{A}+\vec{\nabla} f
\eeq
\beq
\phi\rightarrow\phi-\frac{1}{c}\frac{\partial f}{\partial t}
\eeq
\emath
leaves the electric and magnetic fields unchanged. 

The second London equation
Eq. (1b) can be written as
\beq
\vec{J}=-\frac{n_s e^2}{m_e c}\vec{A}
\eeq
However the right-hand-side of  this equation is not gauge-invariant, while the left-hand side is.
From the continuity equation, 
\beq
\vec{\nabla}\cdot\vec{J}+\frac{\partial \rho}{\partial t}=0
\eeq
it can be seen that Eq. (6) is only valid with a choice of gauge that satisfies
\bmath
\beq
\vec{\nabla}\cdot \vec{A}=\frac{m_e c}{n_se^2}\frac{\partial \rho}{\partial t}  
\eeq 
In particular, in a time-independent situation
(i.e. when electric and magnetic fields are time-independent)
 the vector potential   $\vec{A}$ in Eq. (6) is necessarily transverse, i.e.
\beq
 \vec{\nabla}\cdot\vec{A}=0
 \eeq
 \emath
It is currently generally accepted that the London Eq. (6) is valid for superconductors in all situations, time-independent or not,
with the gauge chosen so that Eq. (8b) holds (London gauge), and hence $\vec{\nabla}\cdot \vec{J}= 0$ from Eq. (6). 
Together with Eq. (1a) this is equivalent to assuming that no longitudinal electric fields
(i.e. $\vec{\nabla}\cdot \vec{E}\ne 0$) can exist inside superconductors, as can be seen
by applying the divergence operator to Eq. (1a). 

However, the possibility of an electrostatic field inside superconductors was recently suggested by the theory of
hole superconductivity\cite{atom,hole,undressing}, which predicts that negative charge is expelled from the interior of superconductors 
towards the surface. The conventional London electrodynamics is incompatible with that possibility,
however the alternate electrodynamics proposed here is not. We have already discussed some consequences for the
electrostatic case in recent work\cite{charge}.

Recently, Govaerts, Bertrand and Stenuit have discussed an alternate Ginzburg-Landau formulation for superconductors that 
is relativistically covariant and has some common elements with the theory discussed here\cite{newwork}. 
For the case of a uniform order parameter their equations
reduce to the early London theory\cite{london1} that allows for electrostatic fields within a penetration length of the surface of
a superconductor, as our theory also does. However in contrast to the theory discussed here, 
the theory of Gorvaets et al does not allow for electric charge nor electric fields deep in the interior of superconductors.

\section{Difficulties with the London model}

Following London\cite{london2}, let us assume that in addition to superfluid electrons there are
normal electrons, giving rise to a normal current 
\beq
\vec{J}_n=\sigma_n\vec{E}
\eeq
so that the total current $\vec{J}=\vec{J}_n+\vec{J}_s$ satisfies
\beq
\frac {\partial \vec{J}}{\partial t}=\frac{c^2}{4\pi\lambda_L^2}\vec{E} +\sigma_n\frac{\partial \vec{E}}{\partial t}
\eeq
with the London penetration depth $\lambda_L$ given by
\beq
\frac{1}{\lambda_L^2}=\frac{4\pi n_s e^2}{m_e c^2}     .
\eeq
Applying the divergence operator to Eq. (10) and using Eqs. (2c) and (7) leads to 
\beq
\frac{\partial^2\rho}{\partial t^2} +4\pi \sigma_n \frac{\partial \rho}{\partial t}+\frac{c^2}{\lambda_L^2} \rho=0
\eeq
London argued\cite{london2} that Eq. (12) leads to a rapid decay of any charge buildup in superconductors, given by the time scales
\beq
\tau_{1,2}^{-1}=2\pi\sigma_n \pm [(2\pi\sigma_n)^2-(\frac{c}{\lambda_L})^2]^{1/2}
\eeq
the slower of which he estimated to be $10^{-12} sec$, and consequently that any charge buildup inside the
superconductor can be ignored. He concluded from this that it is reasonable to assume $\rho=0$ inside superconductors,
which from the continuity equation and Eq. (6) implies Eq. (8b).
The same argument is given in Ref. \cite{rick}.

However, this numerical estimate is based on assuming for $\sigma_n$ a value appropriate for the normal
state, and for $\lambda_L$ its value near zero temperature. Instead, the temperature dependence of $\sigma_n$ and 
$\lambda_L$ should be considered. The conductivity of normal carriers $\sigma_n$ should be proportional to 
the number of 'normal electrons', that goes to zero as the temperature approaches zero, hence $\sigma_n\rightarrow 0$ as $T\rightarrow 0$.  
On the other hand, the number of superfluid electrons $n_s$ goes to zero as $T$ approaches
$T_c$, hence $\lambda_L\rightarrow \infty$ as $T\rightarrow T_c$.  These facts lead to a strong temperature
dependence of the relaxation times in Eq. (13), and in fact to the conclusion that long-lived charge
fluctuations should exist both when $T\rightarrow T_c$ and when $T\rightarrow 0$.

Eq. (12) describes a  damped harmonic oscillator. The crossover between overdamped (at high $T$) and
underdamped (low $T$) regimes occurs for
\beq
2\pi\sigma_n(T) =\frac{c}{\lambda_L(T)}
\eeq
As $T\rightarrow T_c$, $\lambda_L(T)\rightarrow \infty$ and $\sigma_n(T)$ approaches its normal state value at
$T_c$; as $T\rightarrow 0$, $\lambda_L(T)\rightarrow \lambda_L(0)$ and  $\sigma_n(T)\rightarrow 0$; hence condition Eq. (14) will always be satisfied at
some temperature between $0$ and $T_c$. For example, for a superconductor with  low temperature London penetration depth
$\lambda_L\sim 200 \AA $  and normal state resistivity $\rho\sim 10 \mu \Omega cm$ the crossover would be 
at $T/T_c\sim0.69$;  if $\lambda_L\sim 2000 \AA $  and   $\rho\sim 1000 \mu \Omega cm$,      at $T/T_c\sim0.23$. However, no experimental signature of such a cross-over between overdamped and underdamped charge oscillations at some temperature below $T_c$ has ever
been reported for any superconductor.

For $T$ approaching $T_c$, the slower timescale in Eq. (13) is
\beq
\tau_1\sim \frac{4\pi\sigma_n(T)}{(c/\lambda_L(T))^2}
\eeq
so that for $T$ sufficiently close to $T_c$   overdamped charge fluctuations should
persist for arbitrarily long times. However, such a space charge would give rise to an electric field in the interior
of the superconductor and hence (due to Eq. (1a)) to a current that
grows arbitrarily large, destroying superconductivity.  No  such phenomena  have ever been observed 
in superconductors close to $T_c$ to our knowledge.

At low temperatures,  Eq. (13) implies that underdamped charge oscillations should exist, with damping timescale
\beq
\tau_1 = \frac{1}{2\pi \sigma_n(T)}
\eeq
diverging as $T\rightarrow 0$. The frequency of these oscillations is
\beq
\omega=\sqrt{\frac{c^2}{\lambda_L^2}-4\pi^2\sigma_n^2}
\eeq
and as $T\rightarrow 0$ it approaches the plasma frequency
\beq
\omega_p=\frac{c}{\lambda_L}
\eeq
Hence at zero temperature plasma oscillations in the superconducting state are predicted to exist forever,
just as persistent currents. No such persistent charge oscillations have  been observed to our knowledge.

Reference \cite{rick} recognized this difficulty and suggested that the London electrodynamic equations should  only be assumed to be valid 'in situations in which the fields
do not tend to build up a space charge', hence not in the regions close to $T_c$ and close to $T=0$ discussed above. 
On the other hand, the consequences of London
equations concerning magnetic fields and persistent currents are generally believed to be valid for
arbitrary temperatures, and indeed experiments support this expectation. The fact that the implications of London's equations
concerning the behavior of the charge density in superconductors appear to have at best a limited range of validity is disturbing
and suggests a fundamental inadequacy of these equations.

A related difficulty with London's equations arises from consideration of the equation for the 
electric field. Taking the time derivative of Eq. (2b) and using (2a), (2c) and (1a) yield
\beq
\nabla^2\vec{E}=\frac{1}{\lambda_L^2}\vec{E}+\frac{1}{c^2}\frac{\partial^2\vec{E}}{\partial t^2}+4\pi\vec{\nabla}\rho
\eeq
For slowly varying electric fields and assuming no charge density in the superconductor
\beq
\nabla^2\vec{E}=\frac{1}{\lambda_L^2}\vec{E}
\eeq
which implies that an electric field penetrates a distance $\lambda_L$, as a magnetic field does. Indeed, electromagnetic
waves in superconductors penetrate a distance $\lambda_L$, hence Eq. (20) properly describes the screening of transverse
electric fields. However Eq. (20) does not depend on the frequency and hence should remain valid in the static limit; but such
a situation is incompatible with the first London equation (1a), as it would lead to arbitrarily large currents. This then
suggests that application of an arbitrarily small static or quasi-static electric field should lead to destruction of the
superconducting state, which is not observed experimentally. 
 
In a normal metal these difficulties do not arise. Applying the divergence operator to Eq. (9) and using (2c) and (7) yields
\beq
\frac{\partial \rho}{\partial t}=-4\pi\sigma_n \rho
\eeq
which predicts that any charge fluctuation in the interior of metals is screened over a very short time ($\sim 10^{-17} secs$), and
consequently that a longitudinal electric field cannot penetrate a normal metal. A static uniform electric field can penetrate
a normal metal and it causes a finite current to flow whose magnitude is limited by the normal state resistivity.

In summary, London's equations together with Maxwell's equations lead to unphysical predictions regarding the behavior of
superconductors in connection with space charges and electric fields, and to predictions that appear to be contradicted by experiment. How are these 
difficulties avoided in the conventional London picture? By $postulating$ that $\rho=0$ inside
superconductors, and that applied static electric fields do not penetrate the superconductor\cite{london2,tinkham,rick}. These are postulates that
are completely independent of Eqs. (1) and (2), for which no obvious justification within London's theory exists. 
 
\section{The alternate equations}
The possibility of an alternative to the conventional   London equations is  suggested by the
fact that   taking the time derivative of Eq. (6) and using Eq. (4) leads to
\beq
\frac{\partial\vec{J}_s}{\partial t}=\frac{n_se^2}{m_e}(\vec{E}+\vec{\nabla}\phi)
\eeq
without making any additional assumptions on the gauge of $\vec{A}$.    Clearly, Eq. (22) is consistent with the existence of a static electric field in a
superconductor, deriving from an electrostatic potential $\phi$, which will not generate an electric current, contrary to the prediction
of Eq. (1a). 
How it may be possible for a static electric field to exist in a superconductor without generating
a time-dependent current is discussed in ref. \cite{charge}. Assuming that Eq. (22) and Eq. (1a) are equivalent, as is done in the conventional London theory, 
 is tantamount to making an additional $independent$ assumption, namely that
no longitudinal electric field can exist inside superconductors. 

Consequently it seems natural to abandon Eq. (1a) and explore the consequences of  Eq. (22) in its full generality. Starting from Eq. (22), the second London equation (1b) also follows, taking the curl and setting the time integration constant equal to zero as done by London. 
However  to 
completely specify the problem we need further assumptions. Following the early London work\cite{london1} we  take as fundamental equation
Eq. (6) together with the condition that $\vec{A}$ obeys the Lorenz gauge:\bmath
\beq
\vec{J}_s=-\frac{c}{4\pi\lambda_L^2}\vec{A}
\eeq
\beq
\vec{\nabla}\cdot\vec{A}=-\frac{1}{c}\frac{\partial \phi}{\partial t}
\eeq
\emath
These equations imply that the magnetic vector potential that
enters into London's equation (23a) is transverse in a static situation as in London's case, but has a longitudinal component
in a time-dependent situation.

Application of the divergence operator to Eq. (23a), together with Eq. (23b) and the continuity Eq. (7) then leads to
\beq
\frac{\partial \rho}{\partial t}=-\frac{1}{4\pi\lambda_L^2}\frac{\partial \phi}{\partial t}
\eeq
and integration with respect to time to
\beq
\phi(\vec{r},t)-\phi_0(\vec{r})=-4\pi\lambda_L^2(\rho(\vec{r},t)-\rho_0(\vec{r}))
\eeq
where $\phi_0(\vec{r})$ and $\rho_0(\vec{r})$ are constants of integration. A possible choice would be
$\phi_0=\rho_0=0$. Instead, motivated by the theory of hole superconductivity\cite{hole,undressing,atom, charge}, we choose
\beq
\rho_0(\vec{r})=\rho_0>0
\eeq
that is, a uniform $positive$ constant in the interior of the superconductor. Equation (25) then implies that the electrostatic potential $\phi(\vec{r},t)$
equals $\phi_0(\vec{r})$ when the charge density inside the superconductor is constant, uniform and equal to $\rho_0$, hence from Maxwell's
equations we deduce that $\phi_0(\vec{r})$ is given by
\beq
\phi_0(\vec{r})=\int_V d^3r' \frac{\rho_0}{|\vec{r}-\vec{r}'|}
\eeq
where the integral is over the volume of the superconducting body.

In summary, we propose that the macroscopic electrodynamic behavior  of a superconductor is described by Eqs. (23) 
 and a single positive number $\rho_0$, which together with Eq. (27) determines the integration constants
 in Eq. (25). $\rho_0$ is a function of temperature, the particular material, and the dimensions and shape of the
superconducting body\cite{charge}. In the following we explore some consequences of this proposal.

\section{Electrostatics}

For a static situation Eq. (25) is
\beq
\phi(\vec{r})=\phi_0(\vec{r})-4\pi\lambda_L^2(\rho(\vec{r})-\rho_0)
\eeq
with $\phi_0(\vec{r})$ given by Eq. (27)\cite{note}. Using Poisson's equation we obtain for the charge density inside the
superconductor
\beq
\rho(\vec{r})=\rho_0+\lambda_L^2\nabla^2\rho(\vec{r})
\eeq
Inside and outside the superconductor the electrostatic potential obeys
\bmath \beq
\nabla^2(\phi(\vec{r})-\phi_0(\vec{r}))=\frac{1}{\lambda_L^2}(\phi(\vec{r})-\phi_0(\vec{r}))
\eeq
\beq
\nabla^2\phi(\vec{r})=0
\eeq
\emath
respectively. 
Furthermore we assume that no surface charges can exist in superconductors, hence that both $\phi$ and its normal derivative
$\partial \phi / \partial n$ are continuous across the surface of the superconducting body. For given $\rho_0$, these equations
have a unique solution for each value of the average charge density of the superconductor
\beq
\rho_{ave}=\frac{1}{V}\int d^3r \rho(\vec{r})
\eeq
In particular, if $\rho_{ave}=\rho_0$ the solution is $\rho(\vec{r})=\rho_0$ everywhere inside the superconductor and
 $\phi(\vec{r})=\phi_0(\vec{r})$, with $\phi_0$ given by Eq. (27), valid both
inside and outside the superconductor.

For the general case $\rho_{ave}\neq \rho_0$ the solution to these equations can be obtained numerically for any given body
shape by the procedure discussed in ref. \cite{charge}. For a spherical body an analytic solution exists, and we speculate
 that analytic solutions may exist for other shapes of high symmetry. Quite generally, Eq. (29) implies that for body
dimensions much larger than the penetration depth the charge density is $\rho_0$  deep in the interior of the superconductor,
and the potential   is $\phi_0(\vec{r})$.
Deviations of $\rho(\vec{r})$  from $\rho_0$ exist within a layer of thickness $\lambda_L$ of the surface, to  give rise to the given
$\rho_{ave}$. In particular, for a charge neutral superconductor, $\rho_{ave}=0$, excess negative charge will exist near
the surface as discussed in ref. \cite{charge}.

The electrostatic field is obtained from the usual relation
\beq
\vec{E}(\vec{r})=-\vec{\nabla} \phi(\vec{r})
\eeq
and also satisfies the equation
\beq
\vec{E}(\vec{r})=\vec{E}_0(\vec{r})+\lambda_L^2\nabla^2\vec{E}(\vec{r})
\eeq
with $\vec{E}_0(\vec{r})=-\vec{\nabla}\phi_0(\vec{r})$. Deep in the interior, $\vec{E}(\vec{r})=\vec{E}_0(\vec{r})$.
Because of Eq. (22), no current is generated by this electrostatic field.

If an external electrostatic field is applied, the charge density will rearrange so as to screen the external field over a distance
$\lambda_L$ from the surface. This is easily seen from the superposition principle, since the total electric field will be the sum of the
original field and an added field $\vec{E}'(\vec{r})$ that satisfies
\beq
\vec{E}'(\vec{r})=\lambda_L^2\nabla^2\vec{E}'(\vec{r})
\eeq
inside the superconductor, and approaches the value of the applied external field far from the superconductor. Equation (34) implies
that the additional field $\vec{E}'(\vec{r})$ is screened within a distance $\lambda_L$ from the surface, just as an applied magnetic
field would be screened. Quantitative results for general geometries can be obtained by the same procedure outlined in ref. \cite{charge} and
will be discussed elsewhere.

For the particular case of a spherical geometry the solution to these equations is easily obtained. The electrostatic potential for a sphere of radius
$R$ and total charge $q$ is $\phi(r)=\tilde{\phi}(r)+\phi_0(r)$, with 
\bmath
\beq
\tilde{\phi}(r)=\frac{Q}{f(R/\lambda_L)}\frac{sinh (r/\lambda_L)}{r} \ \ ;\ \ r<R
\eeq
\beq
\tilde{\phi}(r)=\frac{Q}{f(R/\lambda_L)}\frac{sinh (R/\lambda_L)}{R} +Q(\frac{1}{R}-\frac{1}{r}) \ ; \ r>R
\eeq
\emath
and 
\bmath
\beq
Q=Q_0-q
\eeq
\beq
Q_0=\frac{4\pi}{3}R^3\rho_0
\eeq
\beq
f(x)=x cosh x -sinh x
\eeq
\beq
\phi_0(r)=\frac{Q_0}{2R}(3- \frac{r^2}{R^2}) ; r<R
\eeq
\beq
\phi_0(r)=\frac{Q_0}{r} ; r>R
\eeq
\emath
and the electric field and charge density follow from Eqs. (32) and (28).
In the presence of a uniform applied electric field $E_{ext}$ the potential is
$\phi(r)+\phi'(r,\theta)$, with
\bmath
\beq
\phi'(r,\theta)=a\frac{\lambda_L^2}{r^2} f(r/\lambda_L)E_{ext}cos\theta \ \  ; \ \  r<R
\eeq
\beq
\phi'(r,\theta)=(\frac{\alpha}{r^2}-r)E_{ext}cos\theta \ \  ; \ \  r>R
\eeq 
\emath
with
\bmath
\beq
a=-\frac{3R}{sinh(R/\lambda_L)}
\eeq
\beq
\alpha=R^3[1-3\frac{\lambda_L^2}{R^2} \frac{f(R/\lambda_L)}{sinh(R/\lambda_L)}]
\eeq
\emath
The induced charge density is easily obtained from Eq. (28). Note that 
a dipole moment $\alpha E_{ext}$ is induced on the sphere and that the polarizability $\alpha$ becomes increasingly reduced compared
with the normal metal value $\alpha=R^3$ as the ratio of radius to penetration length decreases. 
For $R>>\lambda_L$ eq. (38b) yields $\alpha=(R-\lambda_L)^3$ as one would expect, and for
$R<<\lambda_L$, $\alpha=R^5/15\lambda_L^2$.

\section{Magnetostatics}

The magnetostatics for our case is identical to the conventional case. From Eq. (23a), (2b) and (2d) it follows that 
\beq
\vec{B}(\vec{r})=\lambda_L^2\nabla^2\vec{B}(\vec{r})
\eeq
giving rise to the usual Meissner effect. The generated screening supercurrent is not sensitive to the presence of the
electrostatic field.

\section{Electrodynamics}
Using   Eqs.  (23) and Maxwell's equations the following equations result for the electrodynamic behavior of
superconductors:
\bmath
\beq
\nabla^2\vec{B}=\frac{1}{\lambda_L^2}\vec{B}+\frac{1}{c^2} \frac{\partial^2\vec{B}}{\partial t^2}
\eeq
\beq
\nabla^2(\vec{E}-\vec{E}_0)=\frac{1}{\lambda_L^2}(\vec{E}-\vec{E_0})+\frac{1}{c^2} \frac{\partial^2(\vec{E}-\vec{E}_0)}{\partial t^2}
\eeq
\beq
\nabla^2\vec{J}=\frac{1}{\lambda_L^2}\vec{J}+\frac{1}{c^2} \frac{\partial^2\vec{J}}{\partial t^2}
\eeq
\beq
\nabla^2 (\rho- \rho_0)=\frac{1}{\lambda_L^2}(\rho-\rho_0)+\frac{1}{c^2} \frac{\partial^2(\rho- \rho_0)}{\partial t^2}
\eeq
\emath
so that all quantities obey exactly the same equation. Eqs. (40a,b,c) with $\rho_0$ and $\vec{E}_0=0$ are also obtained from the
conventional London equations only if one imposes the additional assumption that $\rho=0$ inside the superconductor. For example, the
equation for the electric field in London's case is
\beq
\nabla^2\vec{E}=\frac{1}{\lambda_L^2}\vec{E}+\frac{1}{c^2} \frac{\partial^2\vec{E}}{\partial t^2} +4\pi\vec{\nabla{\rho}}
\eeq
instead of Eq. (40b).

The simplicity of   eqs. (40) derives from the fact that the theory is relativistically covariant\cite{london1}. 
This is seen as follows. We define the current 4-vector in the usual way
\beq
{\it{J}}=(\vec{J}(\vec{r},t),ic\rho(\vec{r},t))
\eeq
and the four-vector potential
\beq
{\it{A}}=(\vec{A}(\vec{r},t),i\phi(\vec{r},t))
\eeq
The continuity equation sets the four-dimensional divergence of the four-vector $\it{J}$ equal to zero,
\beq
Div {\it{J}}=0
\eeq
where the fourth derivative is $\partial /\partial (ict)$, and the Lorenz gauge condition Eq. (23b) 
sets the divergence of the four-vector ${\it{A}}$ to zero
\beq
Div {\it{A}}=0
\eeq
Furthermore we define the four-vectors associated with the positive uniform charge density $\rho_0$ and its associated
current $\vec{J}_0$, denoted by ${\it J_0}$, and the associated four-vector potential ${\it A_0}$. In the frame of reference where the
superconducting body is at rest the spatial part of these four-vectors is zero, hence
\bmath
\beq
{\it{J_0}}=(0,ic\rho_0)
\eeq
\beq
{\it{A_0}}=(0,i\phi_0(\vec{r}))
\eeq
\emath
in that reference frame. 
In any inertial reference frame, ${\it A_0}$ and ${\it J_0}$  are obtained by Lorentz-transforming Eq. (46) and are related by
\beq
\Box^2 {\it{A_0}}=-\frac{4\pi}{c}{\it{J_0}}
\eeq
with the d'Alembertian operator
\beq
\Box^2=\nabla^2-\frac{1}{c^2}\frac{\partial^2}{\partial t^2}
\eeq
 according to Maxwell's equations, just as the four-vectors ${\it J}$ and ${\it A}$ obey
\beq
\Box^2 {\it{A}}=-\frac{4\pi}{c}{\it{J}}.
\eeq
Our fundamental equation is
 then the relation between four-vectors
 \bmath
\beq
\Box^2({\it{A}}-{\it{A_0)}}=\frac{1}{\lambda_L^2}({\it{A}}-{\it{A_0}})
\eeq
or, equivalently using Eqs. (47) and (49)
\beq
{\it{J}}-{\it{J_0}}=-\frac{c}{4\pi\lambda_L^2}({\it{A}}-{\it{A_0}})
\eeq
\emath
which we propose to be valid in any inertial reference frame. In the frame of reference at rest with respect to the
superconducting body, ${\it{J_0}}$ and ${\it{A_0}}$ have only time-like components, in another reference frame they will
also have space-like components. The spatial and time-like parts of Eq. (50b) give rise to Eqs. (23a) and (25)
respectively. 

Equations (40) can also be written in covariant form. Eqs. (40c) and (d) are
\bmath
\beq
\Box^2({\it{J}}-{\it{J_0)}}=\frac{1}{\lambda_L^2}({\it{J}}-{\it{J_0}})
\eeq
and Eqs. (40a,b)
\beq
\Box^2 ({\it{F}}- {\it{F_0}})=\frac{1}{\lambda_L^2}({\it{F}}-{\it{F_0}})
\eeq
\emath
where ${\it{F}}$ is the usual electromagnetic field tensor and ${\it{F_0}}$ is the field tensor with entries
$\vec{E}_0$ and $0$ for $\vec{E}$ and $\vec{B}$ respectively when expressed in the reference frame at rest
with respect to the ions.

\section{Relativistic covariance}

The fundamental equation (50) relates the $relative$ $motion$ of the superfluid and the positive background, with four-currents
${\it J}$ and ${\it J_0}$, and associated vector potentials ${\it A}$ and ${\it A_0}$ respectively. It is a covariant relation between
four-vectors. This means it is valid in any inertial reference frame, with the four-vectors in different inertial frames
related by the Lorentz transformation connecting the two frames.

In contrast, the conventional London equation Eq. (6) with the condition Eq (8b) is $not$ relativistically covariant, rather it
is only valid in the reference frame where the superconducting solid is at rest.
It can certainly be argued that
the frame where the superconducting solid is at rest is a preferred reference frame, different from any other reference frame. 
(In fact, our covariant equations (50) do recognize the special status of that reference frame, as the only frame where the spatial part of the four-vector $J_0$ is zero.) Hence it is not  a priori obvious that the electrodynamics of superconductors has to be describable by relativistically
covariant equations. 

However, imagine a superconductor with a flat surface of arbitrarily large extent, and an observer moving parallel to this surface.
Since in the conventional London theory the superconductor is locally charge neutral, the state of motion of the observer relative to the superconductor is determined by its motion relative to lateral surfaces of the
body that are arbitrarily far away. According to the conventional London equations this observer should be able to detect
instantaneously any change in the position of these remote lateral surfaces. Since information cannot travel at speeds larger than the
speed of light, this would imply that London's equations cannot be valid for times shorter than the time a light signal takes
in traveling from the observer to the lateral surface; this time increases as the dimensions of the body increase, and can
become arbitrarily large unless one assumes there is some limiting value to the possible size of a superconductor.
Clearly, to describe the electrodynamics of superconductors with local equations whose range of validity depends on the
dimensions of the body is not satisfactory. Our covariant equations avoid this difficulty.

Note also that a relativistically covariant formulation does not make sense if $\rho_0=0$, as assumed in the
early London work\cite{london1}. In that case, the fundamental equation
$\Box^2 A=-(4\pi/c) J$ makes no reference to the state of motion of the solid and
 would describe the same physics irrespective of the relative motion of the
superfluid and the solid, in contradiction with experiment.

\section{The London moment}

The presence of a magnetic field in rotating superconductors\cite{london2}
 presents another difficulty in the conventional London theory. A rotating superconductor has a magnetic field in its interior given by\cite{londonmoment}
\beq
\vec{B}=-\frac{2 m_e c}{e} \vec{\omega}
\eeq
with $\vec{w}$ the angular velocity. This is explained as follows: in terms of the superfluid
velocity $\vec{v}_s$ the superfluid current is $\vec{J}_s=n_s e \vec{v}_s$, so that Eq. (6) is
\beq
\vec{v}_s=-\frac{e}{m_e c} \vec{A}
\eeq
Next one assumes that in the interior the superfluid rotates together with the lattice, so that at position $\vec{r}$
\beq
\vec{v}_s=\vec{\omega}\times \vec{r}
\eeq
and replacing in Eq. (53) and taking the curl, Eq. (52) results.

The problem with this explanation is that, as discussed earlier, the conventional London equations are not
covariant, rather they are valid only in the rest frame of the superconducting body. However in writing Eq. (54) one
is affirming the validity of London's equation in a frame that is $not$ the rest frame of the solid, but is a 
particular inertial frame. To assume the validity of the equations with respect to one particular inertial frame that is not
the rest frame of the solid but not in other inertial frames does not appear to be logically consistent and is reminiscent 
of the old theories of the 'aether': one is stating that the superconductor 'drags' the 'aether' with it if it translates but not if it 
rotates.

 \section{London rigidity}

In the conventional microscopic theory of superconductivity\cite{tinkham}, as well as in the theory of hole
superconductivitiy\cite{hole}, the superfluid carriers are pairs of electrons with total spin $0$. The Schrodinger equation
for particles of spin $0$ is the non-relativistic limiting form of  the more fundamental  Klein-Gordon equation\cite{baym}. 
It is then reasonable to expect that
the proper microscopic theory to describe superconductivity should be consistent with
Klein-Gordon theory. It is very remarkable that the London brothers in their early work\cite{london1}, without
knowledge of Cooper pairs, suggested the possible relevance of the Klein-Gordon equation to superconductivity.

In Klein-Gordon theory, the components of the current four-vector ${\it{J}}=(\vec{J}(\vec{r},t),ic\rho(\vec{r},t))$ 
 in the presence of the four-vector potential ${\it{A}}=(\vec{A}(\vec{r},t),i\phi(\vec{r},t))$
are given in terms of the scalar wave function $\Psi(\vec{r},t)$ by\cite{baym}
\bmath
\beqn
\vec{J}(\vec{r},t)&=&\frac{e}{2m}[\Psi^*(\frac{\hbar}{i}\vec{\nabla}-\frac{e}{c} \vec{A}(\vec{r},t))\Psi+ \nonumber \\
& &
\Psi(-\frac{\hbar}{i}\vec{\nabla}-\frac{e}{c} \vec{A}(\vec{r},t))\Psi^*]
\eeqn
\beqn
\rho(\vec{r},t)&=&\frac{e}{2mc^2}[\Psi^*(i\hbar \frac{\partial}{\partial t} -e \phi(\vec{r},t))\Psi+ \nonumber \\
& &\Psi(-i\hbar \frac{\partial}{\partial t} -e \phi(\vec{r},t))\Psi^*]
\eeqn
\emath
Deep in the interior of the superconductor we have (for superconductors of dimensions much larger than the penetration
depth) 
\bmath
\beq
\rho(\vec{r},t)=\rho_0
\eeq
\beq
\phi(\vec{r},t)=\phi_0(\vec{r})
\eeq
\beq
\vec{J}(\vec{r},t)=\vec{A}(\vec{r},t)=0
\eeq
\emath
independent of any applied electric or magnetic fields. We now postulate, analogously to the conventional theory\cite{london2,tinkham}, that the wave function
$\psi(\vec{r},t)$ is 'rigid'. In terms of the four-dimensional gradient operator
\beq
Grad=(\vec{\nabla},-\frac{i}{c}\frac{\partial}{\partial t})
\eeq
what we mean is that the combination
\beq
[\Psi^* Grad \Psi-\Psi Grad \Psi^*]
\eeq 
is unaffected by external electric and magnetic fields, as well as by proximity to the boundaries of the superconductor. 
 With $\Psi^*\Psi=n_s$, the superfluid density, this assumption and Eqs. (55, 56) lead to
 \bmath
\beq
\vec{J}(\vec{r},t)=-\frac{n_s e^2}{m_e c} \vec{A}(\vec{r},t)
\eeq
\beq
\rho(\vec{r},t)-\rho_0=-\frac{n_s e^2}{m_e c^2} (\phi(\vec{r},t)-\phi_0(\vec{r}))
\eeq
\emath
i.e. the four components of the fundamental equation Eq. (50b).

We believe this is a compelling argument in favor of the form of the theory proposed here. It is true that for
particles moving at speeds slow compared to the speed of light the Klein-Gordon equation reduces to the
usual Schrodinger equation. However, by the same token the Schrodinger equation satisfied by Cooper pairs
can be viewed as a limiting case of the Klein-Gordon equation. In the conventional theory in the framework of non-relativistic
quantum mechanics, 'rigidity'  of the wave function  leads
to the second London equation. It would be unnatural to assume that the same argument cannot be extended to the
superfluid wavefunction in its relativistic version,  
independent of the speed at which the superfluid electrons are moving.

\section{Dielectric function}

As discussed in previous sections, the electric potential in the interior of the superconductor satisfies
\bmath
\beq
\nabla^2 (\phi- \phi_0) - \frac{1}{c^2} \frac{\partial^2(\phi- \phi_0)}{\partial t^2}=\frac{1}{\lambda_L^2}(\phi-\phi_0)
\eeq
while outside the superconductor the potential satisfies the usual wave equation
\beq
\nabla^2 \phi- \frac{1}{c^2} \frac{\partial^2\phi}{\partial t^2}=0
\eeq
\emath
If a harmonic potential $\phi_{ext}(q,\omega)$ is applied, the superconductor responds with an induced potential
$\phi (q,\omega)$ related to $\phi_{ext}$ by
\beq
\phi (q,\omega)=\frac{\phi_{ext}(q,\omega)}{\epsilon_s(q,\omega)}
\eeq
and we obtain for the longitudinal dielectric function of the superconductor
\beq
\epsilon_s(q,\omega)=\frac{\omega_p^2+c^2q^2-\omega^2}{c^2q^2-\omega^2}
\eeq
with
\beq
\omega_p=\frac{c}{\lambda_L}=(\frac{m_e}{4\pi n_s e^2})^{1/2}
\eeq
the plasma frequency. For comparison, the dielectric function of the normal metal is given by the Linhardt dielectric function\cite{ziman}
\beq
\epsilon_n(q,\omega)=1+\frac{4 \pi e^2}{q^2} \sum_k \frac{f_k-f_{k+q}}{\epsilon_{k+q}-\epsilon_k-\hbar+i\delta}
\eeq
with $f_k$ the Fermi function.

Let us compare the behavior of the dielectric functions for the superconductor and the normal metal. In the static limit we have for the
superconductor from Eq. (62)
\beq
\epsilon_s(q,\omega \rightarrow 0)=1+\frac{\omega_p^2}{c^2 q^2}=1+\frac{1}{\lambda_L^2 q^2}
\eeq
For the normal metal, the zero frequency limit of the Linhardt function yields the Thomas Fermi dielectric function\cite{ziman}
\bmath
\beq
\epsilon_{TF}(q)=1+\frac{1}{\lambda_{TF}^2 q^2}
\eeq
with
\beq
\frac{1}{\lambda_{TF}^2}=4\pi e^2 g(\epsilon_F)
\eeq
\emath
with $g(\epsilon_F)$ the density of states at the Fermi energy. Eqs. (65) and (66) imply that static
external electric fields are screened over distances
$\lambda_L$ and $\lambda_{TF}$ for the superconductor and the normal metal respectively. For free electrons we have,
\beq
g(\epsilon_F)=\frac{3n}{2\epsilon_F}
\eeq
so that
\beq
\frac{1}{\lambda_{TF}^2}=\frac{6\pi  n e^2}{\epsilon_F}=\frac{1}{\lambda_L^2} \times \frac{3 m_ec^2}{2\epsilon_F}
\eeq
assuming the density of superconducting electrons $n_s$ is the same as that of normal electrons.
Eq. (68) shows that the superconductor is much more 'rigid' than the normal metal with respect to charge distortions:
the energy cost involved in creating a charge distortion to screen an applied electric field is $\epsilon_F$ in the normal metal
versus $m_ec^2$ in the superconductor, resulting in the much longer screening length in the superconductor compared to
the normal metal.

The same rigidity is manifest in the dispersion relation for longitudinal charge oscillations. From the zero of the dielectric function Eq. (62) we
obtain for the plasmon dispersion relation in the superconducting state
\beq
\omega_{q,s}^2=\omega_p^2+c^2 q^2
\eeq
Notably, this dispersion relation for longitudinal modes in the superconductor
is identical to the one for transverse electromagnetic waves in this medium.  
 In contrast, the zeros of the Linhardt dielectric function yield for the plasmon dispersion relation\cite{ziman}
\beq
\omega_{q,n}^2=\omega_p^2+\frac{3}{5} v_F^2 q^2
\eeq
so that the plasmon dispersion relation is much steeper for the superconductor, since typically
$v_F  \sim 0.01 c$. We can also write Eqs. (69) and (70) as
\bmath
\beq
\omega_{q,s}=\omega_p(1+\frac{1}{2} \lambda_L^2 q^2)
\eeq
\beq
\omega_{q,n}=\omega_p(1+\frac{9}{10}\lambda_{TF}^2 q^2)
\eeq
\emath
showing that low energy plasmons in superconductors require wavelengths larger than $\lambda_L$ according to the alternate equations, in contrast to normal
metals where the wavelengths are of order $\lambda_{TF}$, i.e. interatomic distances. This again shows the enhanced
'rigidity' of the superconductor with respect to charge fluctuations compared to the normal metal.

\section{Plasmons}

In the conventional London theory one deduces from Eq. (1a) in the absence of normal state carriers
\beq
\frac{\partial^2\rho}{\partial t^2} +\omega_p^2\rho=0
\eeq
upon taking the divergence on both sides and using the continuity equation. The solution of this equation is a charge
oscillation with plasma frequency and arbitrary spatial distribution
\beq
\rho(\vec{r},t)=\rho(\vec{r}) e^{-i\omega_p t}
\eeq
In other words, the plasmon energy is independent of wavevector. This indicates that charge oscillations with
arbitrarily short wavelength can be excited in the superconductor according to London theory. Clearly this is unphysical,
as one would not expect charge oscillations with wavelengths smaller than interelectronic spacings. Consequently
one has to conclude that the 'perfect conductor' equation (1a) necessarily has to break down at sufficiently short lengthscales.

Experiments using EELS (electron energy loss spectroscopy) have been performed on metals in the normal state\cite{eels}
and plasmon peaks have been observed, with plasmon dispersion relation approximately consistent with the
prediction of the Linhardt dielectric function Eq. (70). If London's theory was correct one would expect  that 
in the superconducting state   plasmon excitations energies should be   independent of $q$, at least for values of 
$q^{-1}$ larger than interelectronic distances.

However instead it is expected from BCS theory that plasmons below $T_c$ should be very similar  to plasmons in the normal
state\cite{bcs,rick2}. This expectation is based on the fact that plasmon energies are several orders of magnitude larger than
superconducting energy gaps, and  as a consequence  within BCS theory plasmons should be insensitive to the onset of the superconducting state.
However no EELS experiments appear to have ever been performed on superconducting metals to verify this expectation.

In contrast, the counterpart to Eq. (72) with the alternate equations is the equation for the charge density obtained from (40d):
\beq
\frac{\partial^2 \rho_{pl}}{\partial t^2}+\omega_p^2 \rho_{pl}=c^2\nabla^2 \rho_{pl}
\eeq
where $\rho_{pl}$ is the difference between the charge density and its static value obtained from solution of Eq. (29).   The right-hand side of this equation gives a  'rigidity' to charge oscillations that is
absent in the London model.   From Eq. (74) we obtain the dispersion relation Eq. (68) for plasmons
in the superconducting state.

Furthermore the allowed values of the wavevector $q$ will be strongly constrained in small samples of dimension 
comparable to the penetration depth. Consider for simplicity a small superconducting sphere of radius $R$. A plasma 
oscillation is of the form
\beq
\rho_{pl}(\vec{r},t)=\rho_{pl}\frac{sin qr}{r}e^{i w_{q,s}t}
\eeq
and because of charge neutrality
\beq
\int_Vd^3 r \rho_{pl}(\vec{r},t)=0
\eeq
we obtain the condition on the wavevector
\beq
tan(qR)=qR
\eeq
The smallest wavevector satisfying this condition is
\beq
q=\frac{4.493}{R}
\eeq
so that the smallest frequency plasmon has frequency
\beq
\tilde{\omega}_p=\omega_p\sqrt{1+20.2 \frac{\lambda_L^2}{R^2}}
\eeq
This shift in the plasmon frequency can be very large for small samples. For example, for a sphere of radius
$R=10\lambda_L$, Eq. (74) yields a $20\%$  blue shift in the minimum plasmon frequency.

The optical response of small samples will also be different in our theory. Electromagnetic waves excite surface plasmons
in small metallic particles,
and resonance frequencies depend on sample shape and its polarizability\cite{deheer}. As the simplest example,
for a spherical sample the resonance frequency is given by\cite{deheer}
\beq
\omega_M^2=\frac{Q^2}{M \alpha}
\eeq
with $Q$ the total mobile charge,  $M$ its mass and $\alpha=R^3$ the static polarizability of
 a sphere of radius $R$. We expect the polarizability to become smaller in the
superconducting state as given by eq. (38b), hence our theory predicts an increase in surface plasmon resonance
frequency upon entering the superconducting state for small samples, which should be seen for example in
photoabsorption spectra. For example, for  samples of radius $100 \lambda_L$ and $10 \lambda_L$ the decrease in polarizability predicted by Eq. (38b) is $6\%$ and $27\%$ respectively.
The conventional theory would predict no such change.

In summary, the conventional  theory and our theory lead to very different consequences concerning the
behavior of plasmon excitations when a normal metal is cooled into the superconducting state. In the London theory plasmons
are predicted to be completely dispersionless. Within BCS theory, no change with respect to the normal state is expected either in the plasmon dispersion
relation nor in the long wavelength limit of the plasmon frequency  . 
Instead, in our theory the plasmon dispersion  should be much steeper than in the normal state. Furthermore the minimum
volume plasmon frequency should become larger as the sample becomes smaller, and surface plasmon resonance frequencies
should also become larger for small samples.

\section{Experimental tests}

We do not know of any existing experiments that would be incompatible with the proposed theory. Here we summarize the salient
features of the theory that may be amenable to experimental verification.

\subsection{Screening of applied electric field} Our equations predict that longitudinal electric fields should be screened over distances
$\lambda_L$ rather than the much shorter Thomas Fermi length. This could be tested by measuring changes in capacitance of a 
capacitor with superconducting metal plates, or with a superconductor in the region between plates, upon onset of
superconductivity. Such an experiment was performed by H. London\cite{londonexp} in 1936 but no change was observed.
We are not aware of any follow-up experiment. More accurate experiments should be possible now.

\subsection{Measurement of charge inhomogeneity} The theory predicts excess negative charge within a penetration depth of the
surface of a superconductor and a deficit of negative charge in the interior. It may be possible to detect this charge inhomogeneity
by direct observation, for example by  electron microscopy or  other spectroscopic tools.

\subsection{External electric field} For small superconducting samples of non-spherical shape an electric field is
predicted to exist $outside$ the superconductor near the surface\cite{charge}, which should be detectable by
electrostatic measurements.
Associated with it there should be a force between small superconducting particles leading to the formation of
spherical aggregates.

\subsection{Internal electric field} 

The predicted internal electric field is small on a microscopic scale but extends over macroscopic distances.
Perhaps that makes it experimentally detectable.

\subsection{Plasmons}  Plasmon dispersion relations should be strongly affected by
the transition to superconductivity, with plasmons becoming much stiffer at low temperatures. 
Volume and surface plasmon frequencies should increase in the superconducting state for small samples. 'Small' is defined
by the value of the  ratio of a typical sample dimension to $\lambda_L  $, and effects should be detectable even for this ratio
considerably larger than unity. EELS and optical experiments should be able to detect these changes.  

\subsection{Polarizability}
The polarizability of small samples should be smaller in the superconducting than in the normal state. The effect
should be largest at low temperatures. For samples small compared to the penetration depth the 
polarizability should scale as $V^{5/3}$ rather than $V$, with $V$ the volume.

\section{Discussion}

We have proposed a fundamental reformulation of the conventional London
electrodynamics. 
The proposed theory is relativistically covariant and embodied in the single equation
\beq
\Box^2({\it{A}}-{\it{A_0)}}=\frac{1}{\lambda_L^2}({\it{A}}-{\it{A_0}})
\eeq
with ${\it A}$ the four-vector potential, and ${\it A_0}$ the four-vector potential corresponding to a uniform   charge
density $\rho_0$ at rest in the rest frame of the superconducting body.

Similarly as the conventional London theory\cite{london2}, the equations proposed here can be understood as arising from
the 'rigidity' of the microscopic wave function of the superfluid with respect to perturbations. However, in our case the relevant
microscopic theory is the (relativistically covariant) Klein-Gordon theory, appropriate for spin $0$ Cooper
pairs, rather than the non-relativistic Schrodinger equation. Rigidity in this framework leads inescapably
to the new equation (25) and hence to the four-dimensional
Eq. (81). Furthermore, rigidity in our context refers to both the
effect of external electric and magnetic fields on the superfluid wave function
as well as to the effect on it of proximity to the surface of the superconducting
body in the absence of external fields.

The constant $\rho_0$ may be viewed as a phenomenological parameter arising from integration of equation
(24), independent of any microscopic theory.   Instead, within the theory of 
hole superconductivity $\rho_0$ is  a $positive$ parameter
determined by the microscopic physics\cite{hole,undressing,atom, charge}.
The magnitude of $\rho_0$ does $not$ correspond to an ionic positive charge but is much smaller. It originates in the
absence of a small fraction of conduction electrons from the bulk which, as a consequence of the 
'undressing'\cite{undressing} associated with the transition to superconductivity, have moved outwards to within
a London penetration depth of the surface. In reference \cite{charge} we estimated the excess negative
charge near the surface for Nb to be one extra electron per $500,000$ atoms. For a sample of $1cm$ radius this
correspond to a deficit of 1 electron per $10^{11}$ atoms in the bulk, which gives rise to an electric field
of order $10^6 V/cm$ near the surface. This electric field is very small at a microscopic level, yet it gives rise to
very large potential differences between different points in the interior of a macroscopic sample.

The existence of ${\it A_0}$ in Eq. (81), originating in the positive charge $\rho_0$, breaks charge conjugation
symmetry. As discussed earlier, a non-zero $\rho_0$ is necessary for a meaningful relativistically covariant theory.
The prediction that $\rho_0$ is $positive$ for all superconductors follows from the fundamental electron-hole asymmetry of
condensed matter that is the focus of the theory of hole superconductivity\cite{hole, undressing, holeelec}. 
The fact that electron-hole asymmetry is a fundamental aspect of superconductivity is already experimentally 
established by the fact that the magnetic field of rotating superconductors always points in direction
$parallel$, never $antiparallel$,  to the mechanical angular momentum\cite{eha}.

The electrodynamic equations proposed here describe only the superfluid electrons. At finite temperatures
below $T_c$ there will also be a normal fluid composed of thermally excited quasiparticles. A two-fluid model
description of the system at finite temperatures should be possible and lead to interesting insights.

The theory discussed here appears to be 'simpler' than the conventional London theory in that it requires
fewer independent assumptions. It is also consistent with the more fundamental
Klein-Gordon theory, while the conventional London theory is not, and it avoids certain difficulties of the conventional
London theory.
  We do not believe it contradicts any known experimental facts, except
for the 1936 experiment by H. London\cite{londonexp} which to our knowledge has never been reproduced.
Also, recent remarkable experiments by Tao and coworkers\cite{tao} indicate that the properties of superconductors
in the presence of strong static or quasistatic electric fields are not well understood. 
The theory  leads to many consequences that are different from the conventional theory and should be 
experimentable testable, as discussed in this paper. It should apply to all superconductors, with the magnitude of the 
charge-conjugation symmetry breaking parameter $\rho_0$ being largest for high temperature
superconductors\cite{charge}. 

Recent experiments indicate that optical properties of certain metals in the visible range are
affected by the onset of superconductivity\cite{marel}. This surprising coupling of low and high energy physics,
unexpected within conventional BCS theory, was predicted by the theory of hole superconductivity\cite{apparent}.
In this paper we find that physical phenomena associated with longitudinal plasma oscillations, also a high energy phenomenon,  should
also be affected by superconductivity.  Further discussion of the consequences of this theory   and
its relation with the microscopic physics   will be given  in 
future work.

 \acknowledgements
 The author is grateful to A.S. Alexandrov and L.J. Sham for stimulating discussions, and to D. Bertrand for calling 
 Ref. 9 to his attention.


\begin{references}
\bibitem{note} Note that we are using a slightly different convention than in
ref. \cite{charge}, the definition of the potential $\phi$ differs by the
constant $4\pi\lambda_L^2\rho_0$.
\end{references}
\end{document}